\documentstyle[aps,12pt,cite]{revtex}

\begin{document}

\title {Nonperturbative contributions in quantum-mechanical models: the
instantonic approach}

\author { J. Casahorr\'an \footnote{email:javierc@posta.unizar.es}}
\address{Departamento de F\'{\i}sica Te\'orica, \\
Universidad de Zaragoza, E-50009 Zaragoza, Spain}

\maketitle

\begin{abstract}

We review the euclidean path-integral formalism in connection with the
one-dimensional non-relativistic particle. The configurations which allow
to construct a semiclassical approximation classify themselves into either
topological (instantons) and non-topological (bounces) solutions. The
quantum amplitudes consist on an exponential associated with the classical
contribution multiplied by the fluctuation factor which is given by a
functional determinant. The eigenfunctions as well as the energy eigenvalues
of the quadratic operators at issue can be written in closed form due to
the shape-invariance property. Accordingly we resort to the zeta-function
method to compute the functional determinants in a systematic way. The
effect of the multi-instantons configurations is also carefully considered.
To illustrate the instanton calculus in a relevant model we go to the
double-well potential. The second popular case is the 
periodic-potential where the initial  levels  split into bands.
The quantum decay rate of the metastable states in a cubic model is evaluated
by means of the bounce-like solution.
 
\end{abstract}

\vfill \eject

\section {Introduction.}

The tunneling through classically forbidden regions represents one of the
most striking phenomenon in quantum theory and therefore plays a central
role in many areas of modern physics. On the other hand, together with the
operator formalism of quantum mechanics we have an equivalent description
by means of path-integrals. In such a case the Schrodinger's equation is
substituted by a global approach where the quantum mechanical time evolution
is analysed in terms of a functional integration. Qualitatively speaking, the
path-integral representation corresponds to a sum over all histories allowed
to the physical system we are dealing with. To be precise, we need to take
into account an imaginary exponential of the classical action associated
with every path which fulfill the appropiate initial and end points conditions.
Of course, the quantum amplitude so built is difficult to handle due to the
oscillating character of the exponential at issue. To avoid problems of this
sort we carry out the change $t \rightarrow -i \tau$ (known in the literature
as the Wick rotation). In doing so, we can take advantage of the euclidean 
version of the path-integral which represents by itself a new tool for describing
relevant aspects of the quantum theory. \par

Almost from the very beginning of the subject a semiclassical treatment of the
tunneling phenomena (ranging from periodic-potentials in quantum mechanics to
Yang-Mills models in field theories) has been performed by means of the
so-called instantons. Going to more physical terms, the instantons represent
localised finite-action solutions of the euclidean equation of motion. To be
specific, the euclidean equation is the same as the usual one for our particle
in real time  except that the potential is turned upside down. Although more
massive than the perturbative excitations, the instantons themselves become
stable since an infinite barrier separates them from the ordinary sector of
the model. The stability is reinforced by the existence of a topological
conserved charge which does not arise by Noether's theorem in terms of a well-behaved
symmetry, but characterizes the global behaviour of the system when the
imaginary time is large enough. Accordingly, it comes by no surprise that these
classical solutions have been considered in the literature under the name of
topological configurations. 
Once the appropiate classical solution is well-known, we make an expansion 
around the topological background to evaluate the quadratic fluctuations which
arise in terms of the functional determinant of a second order differential
operator. The integration is solved within the gaussian scheme except for
the zero-modes which appear due to the invariances of the model. To deal with 
these excitations we introduce collective coordinates so that ultimately the
gaussian integration is carried out only along the directions orthogonal
to the zero-modes. \par

As a functional determinant includes an infinite product
of eigenvalues, the result should be in principle a highly divergent expression.
Fortunately we can regularize the fluctuation factors by means of the ratio
of determinants. The eigenfunctions as well as the energy eigenvalues of the
quadratic operators which appear in this article are obtained in closed form
by virtue of the shape-invariance property just derived in one-dimensional
supersymmetric quantum mechanics. Once we reach this point, the evaluation
of the quotient of determinants is borne out by the zeta-function method. \par

On the other hand the use of metastable states, defined as long-lived but
eventually decaying states, is a fact of crucial relevance in the most diverse
branches of physics. The classically stable configurations which allows the
system to decay through quantum tunneling into the true vacuum receives in the
literature the name of false vacuum. \par

For instance, particular interest has been
devoted in recent years to the study of cosmological models where the metastability
properties are of great importance to explain the different steps of the
evolution of the universe as a whole. As regards the very early stages of our
world, it is customarily assumed the existence of an inflationary era in which the
energy density is highly dominated by that of the Higgs field trapped precisely
in a false vacuum. As expected, the end of the inflation period occurs once the
metastable state decays to the true ground-state of the model. \par

Restricting ourselves to quantum mechanics, let us consider a particle moving
under the action of a potential which exhibits a metastable well around the
origin. In such a case, the particle will stay there forever as far as classical
mechanics concerns. Because of tunneling however, the potential allows the
escape of the particle from the single well to infinity. In other words, the
state situated around the origin is quantum mechanically unstable and the 
probability current outside the barrier is non-vanishing. Taking into the
account this fact, the energy eigenvalues become complex so that the imaginary
part itself represents a measure of the decay width. \par

In order to compute the
lifetime of the particle trapped within the single trough of the potential
different methods have been proposed. Again the semiclassical expansion
based on the euclidean version of the path-integral represents the most suitable
approach in this context. When constructing a semiclassical theory for the
decay of metastable states, the euclidean non-topological configuration which
leads the process is often referred to as a bounce solution. To be precise, the
spectrum of small fluctuations around the bounce itself contains a negative
eigenvalue which becomes ultimately responsible for the metastability of the
system as a whole. In summary, both topological (instantons) and non-topological
(bounces) euclidean configurations represent probably the most
adequate tools to study tunneling phenomena. \par

It remains to identify the general class of potentials for which the
functional determinant can be computed according to the method based on
the zeta-function. Our description takes over the shape-invariance properties
of the one-dimensional Schrodinger equation. Among other things, the
shape-invariance represents a sufficient condition for the solvability of the
Schrodinger equation at issue \cite{ju}. In such a case the energy eigenvalues
as well as the eigenfunctions can be obtained in closed form so that the
explicit evaluation of the associated zeta-function is feasible. The curious
reader can find in \cite{ju} a complete list of solvable potentials giving rise
to shape-invariance properties. To be more specific, the cases considered
in this article belong to a general class of Posch-Teller potentials labeled
by $\ell$ ($\ell = 1, 2, ...$). For the sake of shortness only the first
members of the series are discussed in detail although the potentials
associated with higher values of $\ell$ are relevant from a physical 
point of view \cite{ro}. \par

On the other hand, it may be interesting at this point to compare the
results obtained according to the one-loop approximation of the path
integral and the ones derived from the standard WKB method. Almost from
the very beginning of the subject the discussions have been based on the
double-well potential and the periodic sine-Gordon model. However, the
exact spectrum of the Lam\'e potential has been found recently \cite{du}.
Although it is not apparent a priori how to handle more complicated situations
we can conclude that the aforementioned case represents an excellent benchmark
to test the standard WKB procedure. In order not to clutter the paper we
refer to the article of M\"uller-Kirsten et al. \cite{mu} where the
interested reader could find a careful discussion. For background we simply
point out that only the so-called WKB-related method of matched asymptotic 
expansions yields the exact instanton results. In other words, we need to go
beyond the well-known linear connection relations to reach the correct values.
To sum up, it may be worth spelling out how the one-loop approximation
based on the instantonic approach provides a relevant framework
for evaluating physical
magnitudes in a more reliable context than the simple WKB method. \par

The arrangement of the article is as follows. First of all, we review in detail
the main features concerning the instantonic approach in one-dimensional particle
mechanics. To illustrate the method we resort to the double-well potential. In
addition, we also study the harmonic oscillator since represents the reference
for all the models discussed along our analysis. The quantum effects of the
one-instanton as well as the multi-instantons configurations are included
in this chapter. \par

The following section is devoted to the periodic-potential. 
Among other things, the lowest band arising from the splitting of the initial
ground-state is explained by means of the multi-instantons contribution. We
conclude with the consideration of a cubic model to stick out the existence
of metastable states. Different mathematical results concerning one-dimensional
supersymmetric quantum mechanics, shape-invariance properties, zeta-function 
regularization procedures and spectral densities can be consulted in the 
appendices. \par

\section{The instantonic approach.}

To start from scratch let us describe in detail the instanton calculus for
the one-dimensional spinless particle as can be found for instance in
\cite{kl}. The interested reader finds there a comprehensive description
of the whole subject. In order to be specific, we assume that the particle
moves under the action of a confining potential $V(x)$ which yields at
quantum level a pure discrete spectrum of energy eigenvalues. Unless
otherwise noted we choose the origin of the energy so that the minima
of the potential satisfy  $V(x) = 0$. If we set the mass of the particle
equal to unity for notational simplicity, the lagrangian $L$ which governs
the behaviour of the model is given by

\begin{equation}
L = {{1} \over {2}} \left( {{dx} \over {dt}} \right)^2 - V(x)
\label{eq:1}
\end{equation}

Now we can start to quantize the theory. 
If the particle is located at the initial time $t_i = -T/2$ at the point
$x_i$ while one finds it when $t_f = T/2$ at the point $x_f$, the
functional version of the non-relativistic quantum mechanics allows us to
express the transition amplitude in terms of a sum over all paths joining
the world points with coordinates $(-T/2, x_i)$ and $(T/2, x_f)$. 
At this point it proves
convenient to write the action $S$ starting from the lagrangian $L$, i.e.

\begin{equation}
S = \int_{-T/2}^{T/2} L(x,\dot{x}) \ dt
\label{eq:2}
\end{equation}

\noindent so that the contribution of a path has a phase proportional to
the action $S$ itself. Therefore we reduce our problem to the study of
a transition amplitude expressed as

\begin{equation}
<x_f\vert \exp(-i H T) \vert x_i> = N(T) \int [dx] \exp i S[x(t)]
\label{eq:3}
\end{equation}

It may be worth spelling out that $H$ represents the  hamiltonian of the model
at issue while the symbol $[dx]$  indicates the integration over all
functions which fulfill the adequate boundary conditions. The factor
$N(T)$ will be chosen to normalize the amplitude conveniently when we
discuss more mathematically the meaning of (\ref{eq:3}).
As the hamiltonian $H$ gives rise to a pure discrete spectrum of energy
eigenvalues, namely

\begin{equation}
H \vert n > = E_n \vert n >
\label{eq:4}
\end{equation}

\noindent we can write that

\begin{equation}
<x_f\vert \exp(-i H T) \vert x_i> = \sum_{n} \exp(-i E_n T) <x_f \vert n >
<n\vert x_i >
\label{eq:5}
\end{equation}

As we are mainly interested in the first eigenfunctions
of $H$, it proves convenient to transform the exponentials of (\ref{eq:3})
into decreasing exponentials. For such a purpose, we make the change
$t \rightarrow - i \tau$, known in the
literature as the Wick rotation, so that

\begin{equation}
i S[x(t)] \rightarrow \int_{-T/2}^{T/2} \left[ - {{1} \over {2}} 
\left( {{dx} \over {d\tau}} \right)^2 - V(x) \right]  \ d\tau
\label{eq:6}
\end{equation}

In the following we resort systematically to the euclidean action $S_{e}$, i.e.

\begin{equation}
S_e =  \int_{-T/2}^{T/2} \left[  {{1} \over {2}} 
\left( {{dx} \over {d\tau}} \right)^2 + V(x) \right]  \ d\tau
\label{eq:7}
\end{equation}

To sum up, the euclidean formulation of the path-integral corresponds to

\begin{equation}
<x_f\vert \exp(- H T) \vert x_i> = N(T) \int [dx]  \ \exp - S_e[x(\tau)]
\label{eq:8}
\end{equation}

To expose the main features of the semiclassical approximation we start by
considering a particular trajectory $x_{c}(\tau)$ which satisfies the
boundary conditions at issue. Next we perform the expansion of a generic
$x(\tau)$ with identical Dirichlet conditions as $x_{c}(\tau)$ according to

\begin{equation}
x(\tau) = x_c(\tau) + \sum_j c_j \  x_j(\tau)
\label{eq:9}
\end{equation}

\noindent where $x_j(\tau)$ stand for a complete set of orthonormal functions, i.e.

\begin{equation}
\int_{-T/2}^{T/2} x_j(\tau) \  x_k(\tau) \ d\tau = \delta_{jk}
\label{eq:10}
\end{equation}

\noindent  vanishing at our boundary

\begin{equation}
x_j(\pm T/2) = 0
\label{eq:11}
\end{equation}

 We carry things further and make the choice

\begin{equation}
[dx] = \prod_j {{dc_j} \over {\sqrt{2 \pi}}}
\label{eq:12}
\end{equation}

\noindent so that as a matter of fact each path is completely characterized
by the $c_{j}$ themselves. It seems plausible to explain the functional
formalism in terms of the integration over the Fourier coefficients $c_{j}$.
Now we need to identify the quadratic differential operator which gives rise
to the basis $x_j(\tau)$. \par

The quasiclassical approximation (or steepest descent method in more
mathematical language) takes for granted that $x_{c}(\tau)$ represents a
stationary point of the euclidean action. As corresponds to an extremal
path $x_{c}(\tau)$ verifies the equation

\begin{equation}
{{d^2x} \over {d\tau^2}} = V^{\prime}(x)
\label{eq:13}
\end{equation}

\noindent where the prime denotes as usual the derivative with respect to the spatial
coordinate. Notice that (\ref{eq:13}) corresponds to the euclidean equation of motion
for the particle once the potential has been turned upside down. To take into
account the quantum fluctuations we perform a functional expansion about 
$x_{c}(\tau)$. As expected the
crucial feature is the analysis of the second variational derivative since the
linear term is absent since  $x_c(\tau)$ represents an extremal path. We find
that

\begin{equation}
S_e[x_c(\tau) + \delta x(\tau)] = S_{eo} + 
\int_{-T/2}^{T/2} \delta x \left[- {{ 1} \over {2}} {{d^2} \over {d\tau^2}} \delta x +
{{ 1} \over {2}} V^{\prime \prime}[x_c(\tau)] \delta x \right] \ d\tau
\label{eq:14}
\end{equation}

\noindent being $S_{eo}$ the classical action associated with the configuration
$x_c(\tau)$. The conventional form of the semiclassical approximation takes over 
a complete set of eigenfunctions (eigenvalues) of the so-called 
stability equation, i.e.

\begin{equation}
-  {{d^2} \over {d\tau^2}} v_j(\tau) +
 V^{\prime \prime}[x_c(\tau)] v_j(\tau) = \epsilon_j v_j(\tau)
\label{eq:15}
\end{equation}

Now the expansion of (\ref{eq:14}) becomes diagonal so that

\begin{equation}
S_e = S_{eo} + {{ 1} \over {2}} \sum_j c_j^2 \epsilon_j 
\label{eq:16}
\end{equation}

\noindent and therefore the euclidean transition amplitude reduces itself to

\begin{equation}
<x_f\vert \exp(- H T) \vert x_i> = N(T) \exp(- S_{eo}) \ \prod_j 
\epsilon_{j}^{-1/2}
\label{eq:17}
\end{equation}

Almost from the very beginning of the subject the product of eigenvalues has
been written as
 
\begin{equation}
\prod_j \epsilon_{j}^{-1/2} = \left\{ Det \left[
-  {{d^2} \over {d\tau^2}}  +
 V^{\prime \prime}[x_c(\tau)]  \right] \right\}^{-1/2}
\label{eq:18}
\end{equation}

\noindent according to a notation which obviously has its origin in 
the finite-dimensional case.
In principle we assume that (\ref{eq:18}) represents a formal expression where
all the eigenvalues $\epsilon_{j}$ are positive. The difficulties associated
with the existence of zero-modes can be avoided by introducing collective
coordinates, not to mention the negative eigenvalue which yields the
metastability phenomenon (more on this later).  \par

Now we  remind the properties of the
euclidean transition amplitude for the well-grounded problem of the harmonic
oscillator since this model is the reference to deal with the quotient of 
functional determinants
when we go to more relevant systems. If we are dealing with

\begin{equation}
V(x) = {{\nu^2 } \over {2}} \  x^2
\label{eq:19}
\end{equation}

\noindent the amplitude between $x_i = 0$ and $x_f = 0$ is given by

\begin{equation}
<x_f = 0\vert \exp(- H_{ho} T) \vert x_i = 0> = N(T) \left\{ Det \left[
-  {{d^2} \over {d\tau^2}}  +
 \nu^2  \right]\right\}^{-1/2}
\label{eq:20}
\end{equation}

The interested reader can find in \cite{ra} the final form of the amplitude, namely

\begin{equation}
<x_f = 0\vert \exp(- H_{ho} T) \vert x_i = 0> = \left({{\nu} \over {\pi}}\right)^{1/2} \
\left(2 \sinh \nu T \right)^{-1/2}
\label{eq:31}
\end{equation}

Next we must carry things further in order to discuss in detail models which
give rise to relevant instantonic configurations far beyond the trivial cases
where $x_c(\tau)$ reduces itself to a constant. (A complete description of
the subject can be found in the book of Rajaraman \cite{rj}). 
To illustrate these ideas in a simple 
context we again resort to the double-well potential given by 

\begin{equation}
V(x) = {{\omega^2 } \over {8}} (x^2 - 1)^2
\label{eq:33}
\end{equation}

The model is particularly
well adapted for such a purpose since at classical level we find two
degenerate minima located at $x_{-} = - 1$ and $x_{+} =  1$ as
expected when considering a potential which enjoys the discrete symmetry
$x \rightarrow - x$. \par
 
The question we wish to address now is the explicit description of the
tunneling phenomenon in the euclidean version of the path-integral.
To make clear the method we turn upside down the potential as required
by the Wick rotation.  From a physical point of view it
seems reasonable to evaluate the transition amplitude between the points
$x_{-} = - 1$ and $x_{+} =  1$. As regards the topological configuration
which represents the basis for a quasiclassical approximation, we look for
a trajectory with initial $x_{i} = - 1$ at $t_{i} = -T/2$ while
$x_{f} =  1$ when $t_{f} = T/2$. In other words, we need to know in closed
form the instanton just interpolating between the two minima of the potential
$V(x)$. It is customarily assumed that $T \rightarrow \infty$ mainly because
the explicit solution of the problem is much more complicated for finite $T$
(more on this point later). \par

With the appropiate boundary conditions we must integrate (\ref{eq:13}) as
far as the double-well concerns. The $x_{c}(\tau)$ corresponds to the motion of
a particle whose conserved euclidean energy $\tilde{E}$ is given by

\begin{equation}
\tilde{E} = {{1} \over {2}} \left( {{dx_{c}} \over {d\tau}} \right)^2 - V(x_{c})
\label{eq:34}
\end{equation}

According to the Dirichlet conditions required by the form of our transition
amplitude we conclude that  $\tilde{E} = 0$. As a matter of fact the particle
leaves the point $x_{i} = - 1$ and takes an infinite time to climb up to the
top of the mountain $x_{f} =  1$. Once the journey starts the particle cannot
return to the initial point or explore unlimited motions towards plus or minus
infinity. The only dynamical possibility is to reach $x_{f} =  1$ asymptotically
as $t_{f} \rightarrow \infty$. \par

The aforementioned mechanical analogy allows us to solve the problem just by
integration of a first-order differential equation instead of resorting to the
conventional euclidean equation of motion written in (\ref{eq:13}). Therefore
we have

\begin{equation}
 {{dx_{c}} \over {d\tau}} = \pm \sqrt{ 2 V(x_{c})}
\label{eq:35}
\end{equation}

\noindent where we recognize the mechanical version of the well-grounded
Bogomol'nyi condition first introduced in the mid-seventies for the analysis
of field theories \cite{bo}. Now we explicitly solve (\ref{eq:35}) by a simple
quadrature so that

\begin{equation}
x_c(\tau) =  \tanh {{\omega (\tau - \tau_c) } \over {2}} 
\label{eq:36}
\end{equation}

\noindent where the parameter $\tau_c$ indicates the point at which the instanton
makes the jump. As expected the antiinstanton is obtained by means of the
transformation $\tau \rightarrow - \tau$ and allows the connection between
$x_{i} =  1$ and $x_{f} = - 1$. Now we must evaluate the action associated with
$x_c(\tau)$ according to (\ref{eq:7}) to get

\begin{equation}
S_{eo} = {{2 \omega} \over {3}} 
\label{eq:37}
\end{equation}

It should be emphasized that in principle we need classical configurations
for which $x = \pm 1$ at large but finite values $\tau = \pm T/2$. However
the instantons that appear in the literature correspond to solutions where
the particle takes itself an infinite time to complete the trajectory. 
Fortunately the difference is so small that can be ignored mainly because 
ultimately we are precisely interested in the asymptotic limit $T \rightarrow
\infty$. As corresponds to a semiclassical approximation we now evaluate the
quadratic corrections over the background provided by the classical configuration.
In other words, we take into account the functional determinant at issue while
the cubic and higher terms in the quantum fluctuations are neglected. First of
all we need an operator of reference for the explicit evaluation of the
determinant. The most natural choice is of course the harmonic oscillator just
located at either of the two minima of $V(x)$, namely 

\begin{equation}
V^{\prime\prime} ( x = \pm 1) = \omega^2
\label{eq:38}
\end{equation}

\noindent so that our description takes over

\begin{eqnarray*}
<x_f = 1\vert \exp(- H T) \vert x_i = - 1> = N(T) \left\{ Det \left[
- {{d^2} \over {d\tau^2}}  +
 \omega^2  \right]\right\}^{-1/2}   
\end{eqnarray*}
\begin{equation}
\left\{{{Det \left[- (d^2/d\tau^2) + V^{\prime \prime}[x_c(\tau)] \right]} \over
{Det \left[- (d^2/d\tau^2) + \omega^2 \right]}}\right\}^{- 1/2} \ 
\exp(-S_{eo})
\label{eq:39}
\end{equation}
 
\noindent where we have multiplied and divided by the determinant of the
aforementioned harmonic oscillator. 
Incorporating the explicit form of $x_c(\tau)$
the stability equation is given by

\begin{equation}
-  {{d^2} \over {d\tau^2}} x_j(\tau) +
 \left[\omega^2 - {{3\omega^2} \over {2 \cosh^2 (\omega \tau/2)}}\right]
 x_j(\tau) = \epsilon_j x_j(\tau)
\label{eq:40}
\end{equation}

\noindent and corresponds to a Schrodinger's equation with exactly solvable 
Posch-Teller potential. A complete description about the eigenfunctions
and the eigenvalues of (\ref{eq:40}) is possible in terms
of  the shape-invariance properties which appear in supersymmetric
quantum mechanics \cite{ju}. 
This modern framework is particularly well adapted to
incorporate
the zeta-function method \cite{ha}. (The technical details of the method 
can be found in appendix A). \par

Among other things one finds a
zero-mode $x_o(\tau)$ which  could jeopardize the physical mea\-ning of the
determinant. However this eigenvalue $\epsilon_o = 0$ comes by no surprise
since it reflects the translational invariance of the system. To be more specific,
there is one direction in the functional space of the second variations which
is incapable of changing the action. One could just as well discover the
existence of a zero-mode starting from (\ref{eq:13}). If we perform an additional
derivative with respect to $\tau$, then

\begin{equation}
x_{o}(\tau) = {{1} \over {\sqrt{S_{eo}}}} {{dx_{c}} \over {d\tau}}
\label{eq:41}
\end{equation}

\noindent is just the solution of (\ref{eq:15}) with $\epsilon_o = 0$. Notice
the normalization of $x_{o}(\tau)$ which is due precisely to the zero 
euclidean energy condition for $x_{c}(\tau)$.
Going back to the functional arguments we can see how the integration over the
Fourier coefficient $c_{o}$ becomes tantamount to the integration which takes
care of the center of the instanton $\tau_{c}$. 
To fix the
jacobian of the transformation  we consider a first change like

\begin{equation}
\Delta x(\tau) = x_{o}(\tau) \  \Delta c_{o}
\label{eq:42}
\end{equation}

According to the general expression written in (\ref{eq:9}) we observe that under
a shift $\Delta \tau_{c}$ the effect corresponds to

\begin{equation}
\Delta x(\tau) = - \sqrt{S_{eo}} \ x_{o}(\tau) \Delta \tau_{c}
\label{eq:43}
\end{equation}

Next the identification between (\ref{eq:42}) and (\ref{eq:43}) yields

\begin{equation}
d c_{o} = \sqrt{S_{eo}} d\tau_{c}
\label{eq:44}
\end{equation}

\noindent where the minus sign disappears since what matters is the modulus
of the jacobian at issue. In doing so we have that

\begin{eqnarray*}
\left\{{{Det \left[- (d^2/d\tau^2) + V^{\prime \prime}[x_c(\tau)] \right]} \over
{Det \left[- (d^2/d\tau^2) + \omega^2 \right]}}\right\}^{- 1/2} = 
 \end{eqnarray*}
\begin{equation}
\left\{{{Det^{\prime} \left[- (d^2/d\tau^2) + V^{\prime \prime}[x_c(\tau)] \right]} \over
{Det \left[- (d^2/d\tau^2) + \omega^2 \right]}}\right\}^{- 1/2} \
\sqrt{{{S_{eo}} \over {2 \pi}}} \ d\tau_{c}
\label{eq:45}
\end{equation}

\noindent where $Det^{\prime}$ stands for the so-called reduced determinant
once the zero-mode has been explicitly removed. \par

Different procedures have been used in the literature to evaluate the typical
ratio of determinants which appears in (\ref{eq:45}). The reasonings become
much more accesible once the system is enclosed in a box of lenght $T$, thus
avoiding the subleties associated with the continuous spectrum. If one imposes
periodic boundary conditions the most relevant physical information derives
from the analysis of the phase shifts \cite{kl}. At last the limit
$T \rightarrow \infty$ is necessary to achieve a meaningful result. However we
prefer from the very beginning to work in open space and compute the
aforementioned ratio of determinants by resorting to the zeta-function
regularization procedure \cite{ja}.  Going back to (\ref{eq:40}) and
(\ref{eq:45}), we perform the change of variable

\begin{equation}
z = {{\omega \tau} \over {2}}
\label{eq:46}
\end{equation}

\noindent to recognize the presence of the ratio of determinants 

\begin{equation}
Q_{2} = {{Det^{\prime} \ O_{2}} \over {Det \ P_{2}}}
\label{eq:47}
\end{equation}

\noindent together with a  factor $\beta$ given by

\begin{equation}
\beta = {{\omega^2} \over {4}}
\label{eq:48}
\end{equation}

As regards the non-zero spectrum of the operator $O_{2}$ we find a discrete
level with $E_{1} = 3$ while the energy of the scattering states corresponds to

\begin{equation}
E_{k} = k^2 + 4
\label{eq:49}
\end{equation}

\noindent for eigenfunctions derived from the plane waves according to

\begin{equation}
\phi_{2,k}(z) = {{A_{2}^{\dagger}(z)} \over {\sqrt{k^2 + 4}}} \
{{A_{1}^{\dagger}(z)} \over {\sqrt{k^2 + 1}}} \ \left[{{\exp(ikz)} \over
{\sqrt{2 \pi}}} \right]
\label{eq:50}
\end{equation}

If we bear in mind the form of the operators $A_{2}^{\dagger}(z)$ and
$A_{1}^{\dagger}(z)$, namely

\begin{equation}
A_{1}^{\dagger} = - {{d} \over {dz}} +  \tanh z
\label{eq:51}
\end{equation}

\begin{equation}
A_{2}^{\dagger} = - {{d} \over {dz}} + 2 \tanh z
\label{eq:52}
\end{equation}

\noindent we can write that

\begin{equation}
\phi_{2,k}(z) = \left(- k^2 + 2 - {{3} \over {\cosh^2 z}} - 3 i k \tanh z\right)
\ {{\exp(ikz)} \over {\sqrt{2 \pi} \  \sqrt{k^2 + 4} \  \sqrt{k^2 + 1}}}
\label{eq:53}
\end{equation}

As a straightforward consequence of (\ref{eq:53}) the regularized spectral density
$\rho_{r}(k)$ of our problem reads
 
\begin{equation}
\rho_{r}(k) = - {{3 (k^2 + 2)} \over {\pi (k^2 + 4) (k^2 + 1)}}
\label{eq:54}
\end{equation}
 
In this scheme the suitable zeta-function $\zeta_r(s)$ to compute the
ratio of determinants of (\ref{eq:45}) can be expressed as

\begin{equation}
\zeta_r(s) = \zeta_{O_{2}}(s) - \zeta_{P_{2}}(s)
\label{eq:55}
\end{equation}

Armed with this information we write

\begin{equation}
\zeta_r(s) = {{1} \over {\Gamma(s)}} \int_0^{\infty} \mu^{s-1}  d\mu  
\left[\exp(- 3 \mu)  - {{3} \over {\pi}}
\int_{-\infty}^{\infty}  {{ (k^2 + 2) \exp[-(k^2 + 4) \mu] } \over
 { (k^2 + 4) (k^2 + 1)}} \  dk \right]
\label{eq:56}
\end{equation}

\noindent which, in turn, gives rise to \cite{gr}

\begin{equation}
\zeta_r(s) = {{1} \over {3^{s}}} - {{3} \over {\pi}}
\int_{-\infty}^{\infty}  {{ (k^2 + 2) } \over
 {  (k^2 + 1) (k^2 + 4)^{s+1}}} \  dk 
\label{eq:57}
\end{equation}

Next it suffices to break the integral of (\ref{eq:57}) into more 
simple components to obtain $\zeta_r(s)$ in terms of Gamma and Hypergeometric
Functions, namely \cite{gr}

\begin{equation}
\zeta_r(s) = {{1} \over {3^{s}}} - {{3} \over {\sqrt{\pi}}} 
{{1} \over {2^{2s+1}}} {{\Gamma \left(s + {{1} \over {2}}\right)} \over {\Gamma(s + 1)}}
- {{3} \over {\sqrt{\pi}}} 
{{1} \over {2^{2s+3}}} {{\Gamma \left(s + {{3} \over {2}}\right)} \over {\Gamma(s + 2)}}
 \ F\left(1, s + {{3} \over {2}}, s + 2, {{3} \over {4}}\right)
\label{eq:58}
\end{equation}

If we incorporate that \cite{ab}

\begin{equation}
F\left(1, {{3} \over {2}}, 2, {{3} \over {4}}\right) = {{8} \over {3}}
\label{eq:59}
\end{equation}

\noindent the evaluation of $\zeta_r(s)$ at $s = 0$ yields

\begin{equation}
\zeta_r(0) = - 1
\label{eq:60}
\end{equation}

On the other hand the interested reader can find in the appendix B the
fundamental steps which allow ultimately the computation of $\zeta_r^{\prime}(0)$.
Next if we simply collect the partial results obtained along the previous
paragraphs the ratio of determinants $R$ which appears in (\ref{eq:45}), i.e.

\begin{equation}
R = {{Det^{\prime} \left[- (d^2/d\tau^2) + V^{\prime \prime}[x_c(\tau)] \right]} \over
{Det \left[- (d^2/d\tau^2) + \omega^2 \right]}}
\label{eq:61}
\end{equation}

\noindent reduces itself to

\begin{equation}
R = {{1} \over {12 \omega^2}}
\label{eq:62}
\end{equation}

Combining now these informations we obtain the transition
amplitude anticipated in (\ref{eq:39}), namely

\begin{eqnarray*}
<x_f = 1\vert \exp(- H T) \vert x_i = - 1> = 
\end{eqnarray*}
\begin{equation}
 \left({{\omega} \over {\pi}}\right)^{1/2} \
\left(2 \sinh \omega T \right)^{-1/2} \  \sqrt{S_{eo}} \  \sqrt{{{6} \over {\pi}}} \ 
\exp(-S_{eo}) \  \omega \  d\tau_{c} 
\label{eq:63}
\end{equation}

Apart from the first factor, which represents the contribution of the
harmonic oscillator of reference, we get a transition amplitude just
depending on the point $\tau_{c}$ at which the instanton makes precisely the jump.
According to the values of the interval $T$ the result seems plausible
whenever

\begin{equation}
\sqrt{S_{eo}} \  \sqrt{{{6} \over {\pi}}} \ 
\exp(-S_{eo}) \ \omega \  T \ll 1
\label{eq:64}
\end{equation}

\noindent a nonsense condition if $T$ is large enough. However in this
regime we can accommodate configurations constructed of instantons and
antiinstantons which mimic the behaviour of a trajectory strictly derived
from the euclidean equation of motion. In doing so we get an additional bonus
since the integration over the centers of the string of instantons and
antiinstantons is carried out in a systematic way. As a matter of fact the
well-known formula for the level splitting of the double-well potential
appears when considering the effect of the multi-instantons.  \par

As all the above calculations were carried out over a single instanton,
it remains to identify the  contributions which take into account the
effect of a string of widely separated 
instantons and antiinstantons along the $\tau$ axis.
It is customarily assumed that these combinations of topological solutions
represent no strong deviations of the trajectories just derived from the
euclidean equation of motion without any kind of approximation. We shall compute
the functional integral by summing over all such configurations, with
$j$ instantons and antiinstantons centered at points
$\tau_1,...,\tau_j$ whenever

\begin{equation}
-{{T} \over {2}} < \tau_1 < ... < \tau_j < {{T} \over {2}}
\label{eq:65}
\end{equation}

Being narrow enough the regions where the instantons (antiinstantons) make the
jump, the action of the
proposed path is almost extremal. We can carry things further and
assume that the action of the string of instantons and antiinstantons is
given by the sum of the $j$ individual actions. This scheme is well-known 
in the literature where it appears with the name of dilute gas approximation
\cite{co}. \par

Now we can evaluate transition amplitudes
with closed paths with $x_i = -1 = x_f$ for instance, so that the action
at issue $S_t$ will be an even multiple of the single instanton action, i.e.

\begin{equation}
S_t = 2j \ S_{eo}
\label{eq:66}
\end{equation}

As expected the  amplitude between $x_i = - 1$ and
$x_i =  1$ incorporates a contribution given by

\begin{equation}
S_t = (2j + 1) \ S_{eo}
\label{eq:67}
\end{equation}

In addition the translational degrees of freedom of the  separated $j$
instantons and antiinstantons yield an integral of the form

\begin{equation}
\int_{-T/2}^{T/2} \omega d\tau_j \
\int_{-T/2}^{\tau_j} \omega d\tau_{j - 1} ... 
\int_{-T/2}^{\tau_2} \omega d\tau_1 = {{(\omega T)^j} \over {j!}} 
\label{eq:68}
\end{equation}

As regards the quadratic fluctuations around the $j$ topological solutions we have
now that the single ratio of determinants transforms into \cite{co}

\begin{eqnarray*}
\left({{\omega} \over {\pi}}\right)^{1/2} \
\left(2 \sinh \omega T \right)^{-1/2} \ 
\left\{{{Det^{\prime} \left[- (d^2/d\tau^2) + V^{\prime \prime}[x_c(\tau)] \right]} \over
{Det \left[- (d^2/d\tau^2) + \omega^2 \right]}}\right\}^{- 1/2} \longrightarrow
\end{eqnarray*}
\begin{equation}
\left({{\omega} \over {\pi}}  \right)^{1/2} \exp(- \omega T/2) \ 
\left[\left\{{{Det^{\prime} \left[- (d^2/d\tau^2) + V^{\prime \prime}[x_c(\tau)] \right]} \over
{Det \left[- (d^2/d\tau^2) + \omega^2 \right]}}\right\}^{- 1/2}\right]^j 
\label{eq:69}
\end{equation}

\noindent according to the limit of the factor associated with the harmonic
oscillator when $T$ is large. Next we can write the
complete transition amplitudes for the double-well potential so that

\begin{equation}
<x_f = 1\vert \exp(- H T) \vert x_i = - 1> = 
\left({{\omega} \over {\pi}}  \right)^{1/2} \exp(- \omega T/2)
\sum_{j=1}^{\infty} {{(\omega T d)^{2j+1}} \over {(2j+1)!}}
\label{eq:70}
\end{equation}

\noindent where $d$ stands for the so-called instanton density, i.e.

\begin{equation}
d = \sqrt{{{6} \over {\pi}}} \  \sqrt{S_{eo}} \  \exp(-S_{eo})
\label{eq:71}
\end{equation}

In summary

\begin{equation}
<x_f = 1\vert \exp(- H T) \vert x_i = - 1> = 
\left({{\omega} \over {\pi}}  \right)^{1/2} \exp(- \omega T/2) \
\sinh (\omega T d)
\label{eq:72}
\end{equation}

Similarly

\begin{equation}
<x_f = 1\vert \exp(- H T) \vert x_i =  1> = 
\left({{\omega} \over {\pi}}  \right)^{1/2} \exp(- \omega T/2)
\sum_{j=0}^{\infty} {{(\omega T d)^{2j}} \over {(2j)!}}
\label{eq:73}
\end{equation}

\noindent and therefore

\begin{equation}
<x_f = 1\vert \exp(- H T) \vert x_i =  1> = 
\left({{\omega} \over {\pi}}  \right)^{1/2} \exp(- \omega T/2) \
\cosh (\omega T d)
\label{eq:74}
\end{equation}

Resorting to the limit $T \rightarrow \infty$ in (\ref{eq:72}), we obtain the
energy eigenvalues $E_0$ and $E_1$ of the first two levels of
the double-well potential

\begin{equation}
E_0 = {{\omega} \over {2}} - 2 \omega \sqrt{{{\omega} \over {\pi}}} \exp(-2 \omega/3)
\label{eq:75}
\end{equation}

\begin{equation}
E_1 = {{\omega} \over {2}} + 2 \omega \sqrt{{{\omega} \over {\pi}}} \exp(-2 \omega/3)
\label{eq:76}
\end{equation}

As expected the quantum mechanical tunneling transfers the
wave function from one well to the other, thus lifting the degeneracy of the
classical vacua. To close, notice the way in which the energy eigenvalues
depend on the barrier-penetration factor, i.e. the exponential of  minus
the classical action of the instanton at issue. Once we have understood the
main properties of the instanton calculus for the one-dimensional particle
we proceed to extend the method to well different models like a periodic-potential
based on the sine-Gordon theory or the cubic system which
represents by itself an excellent benchmark to discuss the existence of metastable 
states. \par

\section{The periodic-potential.} 

In this section we consider the particle under the effects of a periodic-potential
$V(x)$. At first one can ignore the 
barrier-penetration so that the energy eigenkets are infinitely degenerate. Going
to more physical terms, we construct a set of harmonic oscillators centered
at the bottom of each well. However the quantum-mechanical tunneling changes
dramatically this naif picture. As a matter of fact the single energy eigenvalue
transforms into a continuous band while the eigenstates are given in terms of
Bloch waves. Our objective is to describe how the instanton method serves to
explain these results which are obtained of course by means of conventional 
procedures in any solid-state physics course. To be more specific, the
periodic-potential is the most used model for the study of electrons in a
one-dimensional lattice. \par

Once we have carried out the discussion of the one-instanton amplitude, the
dilute-gas approximation takes over a set of instantons and antiinstantons
freely distributed along the $\tau$-axis. As expected each topological configuration
(instanton or antiinstanton) starts just at the point where its predecessor
ends. In addition, the number of instantons minus the number of
antiinstantons represents the observed change in the coordinate $x$ between
the initial and final points of the transition amplitude at issue. 
For reference we choose a concrete periodic-potential $V(x)$ given by
  
\begin{equation}
V(x) = \omega^4 \left[1 - \cos \left({{x} \over {\omega}}\right) \right]
\label{eq:77}
\end{equation}

\noindent so that the minima of $V(x)$ lie at

\begin{equation}
x = 2 \pi j \omega, \ \ \ \ \ \ \ j \  \epsilon \  Z
\label{eq:78}
\end{equation}

Notice that as usual the minima satisfy $V(x) = 0$. This periodic-potential
can be understood as the quantum-mechanical version of the well-grounded
sine-Gordon model which consists of a real scalar field $\phi(x,t)$ in
$d = 1 + 1$ dimensions governed by a potential function similar to (\ref{eq:77}).
As a matter of fact the instanton we are interested in represents the soliton
in the corresponding bidimensional field theory.  \par

From a qualitative perspective the starting point should be the so-called
{\it tight-binding} approximation where the tunneling of the low-energy
eigenkets from one well to the next one is irrelevant. In doing so we have
an infinitely degenerate ground-state with energy given by

\begin{equation}
E_o = {{V''(2 \pi j \omega)} \over {2}}
\label{eq:79}
\end{equation}

\noindent as corresponds to a set of harmonic oscillators distributed
along the one-dimensional lattice. When considering the tunneling effects
the single level $E_o$ yields a complete band, i.e.

\begin{equation}
E_{\theta} = E_{o} - \alpha \  \cos \theta 
\label{eq:85}
\end{equation}

\noindent being $\alpha$ a constant to determine while $\theta$
serves to denote the different states along the band itself \cite{ki}.  \par

Next we explain the way in which the instanton calculus provides an accurate
description of this Bloch wave. To start from scratch we take into account
the transition amplitude with $x_i = 0$ at $t_i = - T/2$ and $x_f = 2 \pi \omega$
when $t_f =  T/2$. In doing so the functional integral is dominated by the
instanton which interpolates between the two adjoining minima of the potential
$V(x)$. The explicit form of the topological configuration derives once more
from the Bogomol'nyi condition. Solving (\ref{eq:35}) by means of a 
quadrature we have 

\begin{equation}
x_c(\tau) = 4 \omega \arctan \  [\exp\omega (\tau - \tau_c)]
\label{eq:86}
\end{equation}

\noindent so that the action associated with $x_c(\tau)$ reads

\begin{equation}
S_{eo} = 8 \omega^3
\label{eq:87}
\end{equation}

Next we must write the stability equation over the instanton, i.e.

\begin{equation}
-  {{d^2} \over {d\tau^2}} x_j(\tau) +
 \left[\omega^2 - {{2\omega^2} \over {cosh^2 \omega \tau}}\right]
 x_j(\tau) = \epsilon_j x_j(\tau)
\label{eq:88}
\end{equation}

\noindent to obtain again a Posch-Teller potential. In any case we conclude that
  
\begin{eqnarray*}
<x_f = 2 \pi \omega \vert \exp(- H T) \vert x_i = 0> = 
\left({{\omega} \over {\pi}}\right)^{1/2} \
\left(2 \sinh \omega T \right)^{-1/2}
\end{eqnarray*}
\begin{equation}
\left\{{{Det^{\prime} \left[- (d^2/d\tau^2) + V^{\prime \prime}[x_c(\tau)] \right]} \over
{Det \left[- (d^2/d\tau^2) + \omega^2 \right]}}\right\}^{- 1/2} \
\sqrt{{{S_{eo}} \over {2 \pi}}} \ d\tau_{c}
\label{eq:89}
\end{equation}

\noindent since $V''(2 \pi j \omega) = \omega^2$ while $\tau_c$ represents
as usual the collective coordinate of the problem. Now we perform the
change of variable

\begin{equation}
z = \omega \tau
\label{eq:90}
\end{equation}

\noindent so that the relevant quotient of determinants corresponds to 

\begin{equation}
Q_{1} = {{Det^{\prime} \ O_{1}} \over {Det \ P_{1}}}
\label{eq:91}
\end{equation}

\noindent where the global factor is now $\beta = \omega^2$. The discrete
spectrum reduces itself to the zero-mode while the scattering states have
energies of the form

\begin{equation}
E_{k} = k^2 + 1
\label{eq:92}
\end{equation}

\noindent for eigenkets like

\begin{equation}
\phi_{1,k}(z) = 
{{A_{1}^{\dagger}(z)} \over {\sqrt{k^2 + 1}}} \ \left[{{\exp(ikz)} \over
{\sqrt{2 \pi}}} \right]
\label{eq:93}
\end{equation}

In other words

\begin{equation}
\phi_{1,k}(z) = \left(- i k + \tanh z\right)
\ {{\exp(ikz)} \over {\sqrt{2 \pi} \    \sqrt{k^2 + 1}}}
\label{eq:94}
\end{equation}

As regards the expression of the regularized spectral density $\rho_{r}(k)$
we find that

\begin{equation}
\rho_{r}(k) = - {{1} \over {\pi  (k^2 + 1)}}
\label{eq:95}
\end{equation}

The structure that is at work corresponds therefore to the zeta-function
$\zeta_r(s)$ given by

\begin{equation}
\zeta_r(s) = \zeta_{O_{1}}(s) - \zeta_{P_{1}}(s)
\label{eq:96}
\end{equation}

To be more specific

\begin{equation}
\zeta_r(s) = {{1} \over {\Gamma(s)}} \int_0^{\infty} \mu^{s-1}  d\mu  
\left[ - {{1} \over {\pi}}
\int_{-\infty}^{\infty}  {{  \exp[-(k^2 + 1) \mu] } \over
 {  (k^2 + 1)}} \  dk \right]
\label{eq:97}
\end{equation}

In accordance with (\ref{eq:97}) we have that

\begin{equation}
\zeta_r(s) = - {{\Gamma(s + {{1} \over {2}})} \over {\Gamma(s + 1)}}
\label{eq:98}
\end{equation}

Taking advantage of the well-known properties of the Gamma Function we obtain

\begin{equation}
\zeta_r(0) = - 1
\label{eq:99}
\end{equation}

\noindent together with (see appendix B)

\begin{equation}
\zeta'_r(0) = 2 \ln 2
\label{eq:100}
\end{equation}

This approach provides the conventional ratio of determinants, i.e.

\begin{equation}
R = {{1} \over {4 \omega^2}}
\label{eq:101}
\end{equation}

\noindent which in turn gives rise to the transition amplitude we are looking
for. To sum up

\begin{eqnarray*}
<x_f = 2 \pi \omega\vert \exp(- H T) \vert x_i = 0> = 
\end{eqnarray*}
\begin{equation}
\left({{\omega} \over {\pi}}\right)^{1/2} \
\left(2 \sinh \omega T \right)^{-1/2} \  \sqrt{S_{eo}} \  \sqrt{{{2} \over {\pi}}} \ 
\exp(-S_{eo}) \  \omega \  d\tau_{c} 
\label{eq:102}
\end{equation}

Now, when going to the dilute-gas approximation we must consider a set of instantons
and antiinstantons so that each topological configuration starts where its
predecessor ends. As regards the transition amplitude (\ref{eq:102}) we require
that the number of instantons $(n)$ minus the number of antiinstantons $(\bar{n})$
fulfill the condition $n - \bar{n} = 1$ as corresponds to the jump between
$x_i = 0$ and $x_f = 2 \pi \omega$. Because of indistinguishability the dilute-gas
approximation includes a combinatorial factor $F$ in order to eliminate the
over-counting of configurations which correspond to the exchange of center-positions.
For the amplitude (\ref{eq:102}) we find that

\begin{equation}
F = {{(n + \bar{n})!} \over {n! \ \bar{n}!}}
\label{eq:103}
\end{equation}

Notice the difference with the double-well potential where the instantons
strictly alternate with the antiinstantons since the problem has only two
minima so that the combinatorial factor is not necessary at all. \par

By integrating over the translational degrees of freedom we obtain now a
term $I$ of the form

\begin{equation}
I = {{(\omega T)^{n + \bar{n}}} \over {(n + \bar{n})!}}
\label{eq:104}
\end{equation}

Armed with this information we can write the transition amplitude in the
limit $T \rightarrow \infty$, i.e.

\begin{equation}
<x_f = 2 \pi \omega\vert \exp(- H T) \vert x_i = 0> = 
\left({{\omega} \over {\pi}}  \right)^{1/2} \exp(- \omega T/2)
\sum_{n,\bar{n}}^{\infty} {{(\omega T d)^{n + \bar{n}}} \over {n! \ \bar{n}!}}
\delta_{n - \bar{n},1}
\label{eq:105}
\end{equation}

\noindent where $\delta_{n - \bar{n},1}$ is the Kronecker symbol while $d$
stands as usual for the so-called instanton density, namely

\begin{equation}
d = \sqrt{{{2} \over {\pi}}} \  \sqrt{S_{eo}} \  \exp(-S_{eo})
\label{eq:106}
\end{equation}

\noindent with $S_{eo}$ as given in (\ref{eq:87}). Now the sum over $n$ and
$\bar{n}$ decouples by resorting to the integral representation of the
Kronecker symbol, i.e.

\begin{equation}
\delta_{n - \bar{n},1} = \int_0^{2 \pi} \ {{d \theta} \over {2 \pi}} \ 
\exp[ - i \theta (n - \bar{n} - 1)]
\label{eq:107}
\end{equation}

\noindent so that ultimately we have two independent exponential series.
Consequently our main result should be

\begin{eqnarray*}
<x_f = 2 \pi \omega\vert \exp(- H T) \vert x_i = 0> =
\end{eqnarray*}
\begin{equation}
\int_0^{2 \pi} \ {{d \theta} \over {2 \pi}} \exp({i \theta})
\left({{\omega} \over {\pi}}  \right)^{1/2}
\exp(2 \omega d T  \cos \theta  - \omega T/2)
\label{eq:108}
\end{equation}

\noindent so that the left-hand side of (\ref{eq:108}) is dominated when
$T \rightarrow \infty$ by a continuous band parametrised in terms of
$\theta$ with $0 \le \theta \le 2 \pi$. Therefore the energy
of the states contained in this low-lying band should be

\begin{equation}
E_{\theta} = {{\omega} \over {2}} - 2 \omega d \cos \theta
\label{eq:109}
\end{equation}

As expected the result is consistent with the expression written in 
(\ref{eq:85}) so that we have described by means of the instanton
calculus the behaviour of the lowest band of a periodic-potential. \par

\section{Tunneling and decay.}

Next we take into account the decay of metastable states by tunneling processes.
In principle the states with a finite lifetime are analytically described by
means of energy eigenvalues which appear in the lower half of a complex plane
due to a negative imaginary part. According to the temporal evolution of a 
standard wave function it is the case that such imaginary part represents
a measure of the decay width $\Gamma$ through the relation \cite{kl}

\begin{equation}
\Gamma = - 2 Im \  E
\label{eq:110}
\end{equation}

As a matter of fact the negative sign of this imaginary contribution for the energy
eigenvalues serves to prevent the existence of unphysical states with exponentially
growing norm. Our purpose along this section should be the computation of the
magnitude $\Gamma$ for the lowest state of a concrete problem. From a physical point
of view we assume as usual a high barrier-potential so that the lifetime at issue
is long. In other words, the state is approximate stationary so that a quasiclassical
analysis represents the most suitable tool to discuss the main characteristics
of the system. \par

To be more specific, let us consider the point particle in the presence of a cubic
potential like 

\begin{equation}
V(x) = {{\omega^2} \over {2}} x^2- {{1} \over {6}} x^3
\label{eq:111}
\end{equation}

Clearly the potential has a minimum located at $x = 0$. According to classical
mechanics, once the particle is trapped in the single trough of $V(x)$ (with 
negligible kinetic energy) the stability is strictly guaranteed. However the
situation is dramatically different in the quantum regime because of tunneling.
As a matter of fact the particle can go towards $+ \infty$ with a non-vanishing
probability current there. Consequently non-selfadjoint boundary conditions
must be added to the hamiltonian so that the energy eigenvalues become complex. 
The question we wish to address now is the computation of $\Gamma$ for the
lowest state of (\ref{eq:111}). Qualitatively speaking, the classical configuration
which dominates the tunneling process is called a bounce. As expected the
metastability properties derive from the behaviour of the fluctuations built
over the classical configuration. When considering non-topological bounces the
spectrum of the stability equation has one negative eigenvalue which is therefore
responsible for the decay phenomenon. It may be interesting at this point to
remind for instance the case of the double-well potential in  this context. \par

A particle released from rest at the top of the left-hump of $- V(x)$ when
$t_i = - \infty$ arrives at the other hump with velocity zero at time
$t_f = \infty$. In other words, the velocity increases from zero to the 
corresponding maximum and decreases to recover asymptotically the zero
again. But this velocity represents the zero-mode of the stability equation.
Simply put, such zero-mode is a nodeless wave function so that the second
variational derivative of the euclidean action is positive semidefinite. \par

In the semiclassical approximation we shall compute the transition matrix
element between $x_i = 0$ and $x_f = 0$. As usual the development relies on
the solution obtained by solving the imaginary-time equation of motion which
is equivalent to a conventional trajectory in the reversed potential $- V(x)$.
The particle starts at $x_i = 0$, goes through
the minimum of $- V(x)$ and then returns to the initial point. Because of this
come back motion the non-topological solution receives the name of bounce. \par

Next
we proceed in the same manner as in the previous sections. First of all we resort
again to the Bogomol'nyi condition to get the form of the bounce-like 
configuration, i.e.

\begin{equation}
x_c(\tau) = {{3 \omega^2} \over {\cosh^2 [\omega (\tau - \tau_c)/2]}}
\label{eq:112}
\end{equation}

\noindent so that the euclidean action of (\ref{eq:112}) should be

\begin{equation}
S_{eo} = {{ 2 4 \omega^5} \over {5}}
\label{eq:113}
\end{equation}

The Schrodinger equation which governs the behaviour of the fluctuations over
the bounce reads

\begin{equation}
-  {{d^2} \over {d\tau^2}} x_j(\tau) +
 \left[\omega^2 - {{3\omega^2} \over {cosh^2 \omega \tau/2}}\right]
 x_j(\tau) = \epsilon_j x_j(\tau)
\label{eq:114}
\end{equation}

In the case of this cubic potential, the velocity of the particle at issue has
a zero located precisely at the turning point.  To sum up, the second variational derivative of the euclidean action 
is expected to possess one lower even eigenket with negative energy eigenvalue.
This is the point at which the path integral formalism introduces the
metastability. However the explicit derivation of $Im \ E_0$ requires a
careful procedure. \par

First of all we need to perform the integration over the eigenstate $x_{-1}(\tau)$
with eigenvalue $\epsilon_{-1} < 0$. For such a purpose one takes into account the
analytic continuation procedure through a continuous sequence of paths drawn in
the functional space at issue. In order not to clutter the paper we omit the
mathematical details of the computation to emphasize the physical consequences of
the result. The interested reader can find in \cite{kl} a complete description
of the analytic continuation method as tailored to this specific integral. Therefore
we have that

\begin{equation}
\int \exp(- \epsilon_{-1} c_{-1}^2/2) \  {{dc_{-1}} \over {\sqrt{2 \pi}}} =
{{i} \over {2}} {{1} \over {\sqrt{\vert \epsilon_{-1} \vert}}}
\label{eq:115}
\end{equation}

\noindent so that the tunneling rate formula will include the presence of the
additional factor $1/2$ together with the modulus of the quotient of
determinants. According to these arguments the transition amplitude over the
bounce reads

\begin{eqnarray*}
<x_f = 0 \vert \exp(- H T) \vert x_i = 0> = 
\left({{\omega} \over {\pi}}\right)^{1/2} \
\left(2 \sinh \omega T \right)^{-1/2}
\end{eqnarray*}
\begin{equation}
{{i} \over {2}}
\left\vert{{Det^{\prime} \left[- (d^2/d\tau^2) + V^{\prime \prime}[x_c(\tau)] \right]} \over
{Det \left[- (d^2/d\tau^2) + \omega^2 \right]}}\right\vert^{- 1/2} \
\sqrt{{{S_{eo}} \over {2 \pi}}} \ d\tau_{c}
\label{eq:116}
\end{equation}

\noindent because $V''(0) = \omega^2$. Once we incorporate the change of
variable (\ref{eq:46}) the ratio of determinants now corresponds to

\begin{equation}
Q_{3} = \left\vert{{Det^{\prime} \ (O_{3} - 5)} \over {Det \ P_{2}}} \right\vert
\label{eq:117}
\end{equation}

\noindent with the factor $\beta$ given in (\ref{eq:48}). As regards the relevant
discrete spectrum of $O_{3} - 5$ we have $E_{-1} = - 5$ and $E_{1} = 3$ while
the energy of the scattering states corresponds again to (\ref{eq:49}). As a matter
of fact these eigenkets derive from the plane waves, i.e.

\begin{equation}
\phi_{3,k}(z) = {{A_{3}^{\dagger}(z)} \over {\sqrt{k^2 + 9}}}
{{A_{2}^{\dagger}(z)} \over {\sqrt{k^2 + 4}}} \
{{A_{1}^{\dagger}(z)} \over {\sqrt{k^2 + 1}}} \ \left[{{\exp(ikz)} \over
{\sqrt{2 \pi}}} \right]
\label{eq:118}
\end{equation}

\noindent to give

\begin{eqnarray*}
\phi_{3,k}(z) = \left[ - {{15 \sinh z} \over {\cosh^3 z}} - 6 (k^2 - 1) \tanh z +
i k \left({{15} \over {\cosh^2 z}} + k^2 - 11 \right) \right]
\end{eqnarray*}
\begin{equation}
{{\exp(ikz)} \over {\sqrt{2 \pi} \ \sqrt{k^2 + 9} \ \sqrt{k^2 + 4} \  \sqrt{k^2 + 1}}}
\label{eq:119}
\end{equation}

Consequently the regularized spectral density $\rho_{r}(k)$ is of the form

\begin{equation}
\rho_{r}(k) = - {{6 (k^4 + 8 k^2 + 11)} \over {\pi (k^2 + 9) (k^2 + 4) (k^2 + 1)}}
\label{eq:120}
\end{equation}

As expected the zeta-function $\zeta_r(s)$ reads 

\begin{equation}
\zeta_r(s) = \zeta_{O_{3} - 5}(s) - \zeta_{P_{2}}(s)
\label{eq:121}
\end{equation}

\noindent so that

\begin{eqnarray*}
\zeta_r(s) = {{1} \over {\Gamma(s)}} \int_0^{\infty} \mu^{s-1}  d\mu
\end{eqnarray*}
\begin{equation}  
\left[\exp(- 3 \mu) + \exp(- 5 \mu) - {{6} \over {\pi}}
\int_{-\infty}^{\infty}  {{ (k^4 + 8 k^2 + 11) \exp[-(k^2 + 4) \mu] } \over
 {(k^2 + 9) (k^2 + 4) (k^2 + 1)}} \  dk \right]
\label{eq:122}
\end{equation}

In other words

\begin{equation}
\zeta_r(s) = {{1} \over {3^{s}}} + {{1} \over {5^{s}}} - {{6} \over {\pi}}
\int_{-\infty}^{\infty}  {{ (k^4 + 8 k^2 + 11) } \over
 { (k^2 + 9) (k^2 + 1) (k^2 + 4)^{s+1}}} \  dk 
\label{eq:123}
\end{equation}

Breaking as usual (\ref{eq:123}) into more simple integrals we find

\begin{eqnarray*}
\zeta_r(s) = {{1} \over {3^{s}}} + {{1} \over {5^{s}}} - {{3} \over {\sqrt{\pi}}} 
{{1} \over {2^{2s}}} {{\Gamma \left(s + {{1} \over {2}}\right)} \over {\Gamma(s + 1)}}
- {{3} \over {\sqrt{\pi}}} 
{{1} \over {2^{2s+3}}} {{\Gamma \left(s + {{3} \over {2}}\right)} \over {\Gamma(s + 2)}}
 \ F\left(s + {{3} \over {2}}, 1, s + 2, {{3} \over {4}}\right) +
\end{eqnarray*}
\begin{equation}
 + {{5} \over {\sqrt{\pi}}} 
{{1} \over {3^{2s+2}}} {{\Gamma \left(s + {{3} \over {2}}\right)} \over {\Gamma(s + 2)}}
 \ F\left(s + {{3} \over {2}}, s + 1, s + 2, {{5} \over {9}}\right)
\label{eq:124}
\end{equation}

If we bear in mind that

\begin{equation}
F\left({{3} \over {2}}, 1, 2, {{5} \over {9}}\right) = {{9} \over {5}}
\label{eq:125}
\end{equation}

\noindent we get again that

\begin{equation}
\zeta_r(0) = - 1
\label{eq:126}
\end{equation}

Taking advantage of the explicit results collected in appendix B the transition
amplitude at issue reduces itself to

\begin{eqnarray*}
<x_f = 0 \vert \exp(- H T) \vert x_i = 0> = 
\left({{\omega} \over {\pi}}\right)^{1/2} \
\left(2 \sinh \omega T \right)^{-1/2}
\end{eqnarray*}
\begin{equation}
{{i} \over {2}}
\sqrt{S_{eo}} \  \sqrt{{{30} \over {\pi}}} \ 
\exp(-S_{eo}) \  \omega \  d\tau_{c} 
\label{eq:127}
\end{equation}

To close the section we evaluate the transition amplitude by summing over
configurations which contain a set of separated bounces. In this case we have
no restriction to an even or odd number of bounces so that the complete
exponential series arises. To sum up

\begin{equation} 
<x_f = 0 \vert \exp(- H T) \vert x_i = 0> = 
\left({{\omega} \over {\pi}}\right)^{1/2} \
\exp(- \omega T /2) \ \exp(i d \omega/2)
\label{eq:128}
\end{equation}

\noindent for 

\begin{equation}
d = \sqrt{{{30} \over {\pi}}} \  \sqrt{S_{eo}} \  \exp(-S_{eo})
\label{eq:129}
\end{equation}

In doing so we obtain the final expression of the decay width, i.e.

\begin{equation}
\Gamma = \omega  \sqrt{{{30} \over {\pi}}} \  \sqrt{S_{eo}} \  \exp(-S_{eo})
\label{eq:130}
\end{equation}

\noindent where the prefactor in (\ref{eq:130}) is known as the 
{\it quantum attempt frequency} while the second term is interpreted as the
Boltzman weight for the appearance of the bounce which ultimately leads the
tunneling phenomenon.

\vfill \eject

\section {Appendix A.}

In this article we focus the attention on the set of
hamiltonians $O_{\ell}$ given by

\begin{equation}
O_{\ell} = - {{d^2} \over {dz^2}} - {{\ell(\ell+1)} \over {\cosh^{2} z}} + \ell^2
\label{eq:329}
\end{equation}

\noindent where $\ell =1, 2, ...$, although only the first members of the series
lead to relevant models  in physics.
The operator $O_{\ell}$ can be factorized in terms
of a superpotential $W(z,\ell)$ like \cite{ju}

\begin{equation}
W(z,\ell) = \ell \tanh z
\label{eq:330}
\end{equation}

\noindent so that $O_{\ell} = Q_{\ell}^{\dagger}Q_{\ell}$  for

\begin{equation}
Q_{\ell} = {{d} \over {dz}} + \ell \tanh z
\label{eq:331}
\end{equation}

\begin{equation}
Q_{\ell}^{\dagger} = - {{d} \over {dz}} + \ell \tanh z
\label{eq:332}
\end{equation}

The shape-invariance condition appears once we write the partner hamiltonian
$\tilde{O}_{\ell}$ so that in our case the mapping between the old
parameter $\ell$ and the new one $\tilde{\ell}$ reduces to

\begin{equation}
\tilde{\ell} = \ell - 1
\label{eq:333}
\end{equation}

By iteration of the procedure we construct a family of well-behaved superpotentials 
which allow us to solve $O_{\ell}$. The discrete spectrum
includes as usual a normalizable zero-energy mode $\phi_{\ell,o}(z)$ of the form

\begin{equation}
\phi_{\ell,o}(z) = {{\sqrt{2(2\ell - 1)!}} \over {2^{\ell} (\ell - 1)!}} \
{{1} \over {\cosh^{\ell} z}}
\label{eq:334}
\end{equation}

\noindent together with the set of states $\phi_{\ell,m}(z)$ given by

\begin{equation}
\phi_{\ell,m}(z) = {{\sqrt{2(2\ell - 2 m - 1)!}} \over {2^{\ell - m} 
(\ell - m -  1)!}} \ {{1} \over {\sqrt{\prod _{j = 0}^{m - 1} (E_{m} - E_{j})}}} 
\ Q_{\ell}^{\dagger}(z) \ .\ .\ . \
Q_{\ell - m + 1}^{\dagger}(z) \ \left[{{1} \over {\cosh^{\ell - m} z}}\right]
\label{eq:335}
\end{equation}

\noindent with energies

\begin{equation}
E_{m} = \ell^{2} - (\ell - m)^2
\label{eq:336}
\end{equation}

\noindent for $m = 1, ... , \ell - 1$. The continuous spectrum 
corresponds to

\begin{equation}
\phi_{\ell,k}(z) = {{Q_{\ell}^{\dagger}(z)} \over {\sqrt{k^2 + \ell^2}}} \
{{Q_{\ell - 1}^{\dagger}(z)} \over {\sqrt{k^2 + (\ell - 1)^2}}} \ .\ .\ . \
{{Q_{1}^{\dagger}(z)} \over {\sqrt{k^2 + 1}}} \ \left[{{\exp(ikz)} \over
{\sqrt{2 \pi}}} \right]
\label{eq:337}
\end{equation}

\noindent with energy eigenvalues

\begin{equation}
E_{k} = k^2 + \ell^2
\label{eq:338}
\end{equation}

\noindent and  normalization as 

\begin{equation}
\int_{- \infty}^{\infty} \ \phi_{\ell,k}^{*}(z) \ \phi_{\ell,k^{\prime}}(z) \ dz = \delta(k - k^{\prime})
\label{eq:339}
\end{equation}

Let us now
sketch the way in which we can compute the ratio of the determinants associated
with $O_{\ell}$  and $P_{\ell}$ which is given by

\begin{equation}
P_{\ell} = - {{d^2} \over {dz^2}} + \ell^2
\label{eq:350}
\end{equation}

As a matter of fact $P_{\ell}$ itself represents a hamiltonian of comparison for 
$O_{\ell}$. In such a case we need to deal with

\begin{equation}
Q = {{Det^{\prime} \ O_{\ell}} \over {Det \ P_{\ell}}}
\label{eq:351}
\end{equation}

\noindent where $Det^{\prime} \ O_{\ell}$ denotes as usual the reduced determinant
 once the zero-mode has been explicitly removed. To take into account the
continuous spectrum we resort to the density matrix $\Xi_{O_{\ell}}(k,z,w)$
written as \cite{ni}

\begin{equation}
\Xi_{O_{\ell}}(k,z,w) = \phi_{\ell,k}^{*}(z) \ \phi_{\ell,k}(w)
\label{eq:352}
\end{equation}

When going to $P_{\ell}$ we get

\begin{equation}
\Upsilon_{P_{\ell}}(k,z,w) = {{\exp-ik(z - w)} \over {2 \pi}}
\label{eq:353}
\end{equation}

\noindent as expected for a free-particle. The conventional 
subtraction procedure yields the regularized spectral density 
$\rho_r(k)$ expressed as

\begin{equation}
\rho_r(k) = \int_{-\infty}^{\infty} \ \left[\Xi_{O_{\ell}}(k,z,z) - 
\Upsilon_{P_{\ell}}(k,z,z)\right] \ dz
\label{eq:354}
\end{equation}

In this scheme the  zeta-function for the analysis of (\ref{eq:351})
would be \cite{ha}

\begin{equation}
\zeta_r(s) = \zeta_{O_{\ell}}(s) - \zeta_{P_{\ell}}(s)
\label{eq:355}
\end{equation}

\noindent and therefore

\begin{equation}
\zeta_r(s) = {{1} \over {\Gamma(s)}} \int_0^{\infty} \mu^{s-1}  d\mu  
\left\{\exp\left[- \sum_{m = 1}^{\ell - 1} [\ell^2 - (\ell - m)^2] \mu \right] +
\int_{-\infty}^{\infty}  \rho_r(k) \exp\left[-(k^2 + \ell^2) \mu \right] \  dk \right\}
\label{eq:356}
\end{equation}

\vfill \eject

\section {Appendix B.}

Let us come down to the concrete details which allow the evaluation of 
$\zeta_r^{\prime}(0)$ when considering the instanton of the double-well
potential. According to the expression of $\zeta_r(s)$ (see (\ref{eq:58})) we find
that

\begin{equation}
\zeta_r^{\prime}(0) = - \ln 3 + 8 \ln 2 - {{1} \over {2}} - {{3} \over {16}}
F^{\prime}(s + {{3} \over {2}},1,s + 2, {{3} \over {4}})\vert_{s=0} 
\label{eq:357}
\end{equation}

\noindent once we take into account that \cite{ab}

\begin{equation}
\Gamma^{\prime}({{1} \over {2}}) = - \sqrt{\pi} \  (\gamma + 2 \ln 2) 
\label{eq:358}
\end{equation}

\begin{equation}
\Gamma^{\prime}(1) = - \gamma 
\label{eq:359}
\end{equation}

\begin{equation}
\Gamma^{\prime}({{3} \over {2}}) = - \sqrt{\pi} \  ({{\gamma} \over {2}} +  \ln 2 - 1) 
\label{eq:360}
\end{equation}

\begin{equation}
\Gamma^{\prime}(2) = - \gamma + 1
\label{eq:361}
\end{equation}

\noindent where $\gamma$ is the Euler's constant. To obtain the derivative
of the hypergeometric function  we resort to the  integral
representation, namely \cite{ab}

\begin{equation}
F(s + {{3} \over {2}},1,s + 2, {{3} \over {4}}) = {{\Gamma(s+2)} \over
{\Gamma(1) \  \Gamma(s+1)}} \ \int_0^1 (1 - t)^s \ \left(1 - {{3t} \over {4}} 
\right)^{-s-{{3} \over {2}}} \ dt
\label{eq:362}
\end{equation}

\noindent so that

\begin{equation}
F^{\prime}(s + {{3} \over {2}},1,s + 2, {{3} \over {4}})\vert_{s=0} \  = I_1 +
I_2 + I_3
\label{eq:363}
\end{equation}

\noindent for integrals of the form

\begin{equation}
I_1 = \int_0^1 \left(1 - {{3t} \over {4}}\right)^{-{{3} \over {2}}} \ dt
\label{eq:364}
\end{equation}

\begin{equation}
I_2 = \int_0^1   \left(1 - {{3t} \over {4}}\right)^{-{{3} \over {2}}} 
\ \ln(1 - t) \ dt
\label{eq:365}
\end{equation}

\begin{equation}
I_3 = -  \int_0^1   \left(1 - {{3t} \over {4}}\right)^{-{{3} \over {2}}} \ 
\ln\left(1 - {{3t} \over {4}}\right) \ dt
\label{eq:366}
\end{equation}

If we have that 

\begin{equation}
I_1 = {{8} \over {3}}
\label{eq:367}
\end{equation}

\begin{equation}
I_2 = {{32} \over {3}}  \ln {{2} \over {3}}
\label{eq:368}
\end{equation}

\begin{equation}
I_3 = - {{16} \over {3}} + {{32} \over {3}}  \ln 2
\label{eq:369}
\end{equation}

\noindent then $\zeta_r^{\prime}(0)$ is obviously

\begin{equation}
\zeta_r^{\prime}(0) = 4 \ln 2 + \ln 3
\label{eq:370}
\end{equation}

As regards the bounce solution of the cubic potential we obtain that

\begin{equation}
\zeta_r^{\prime}(0) =  9 \ln 2 - \ln 5 + {{1} \over {2}} + {{5} \over {18}}
F^{\prime}(s + {{3} \over {2}},s + 1,s + 2, {{5} \over {9}})\vert_{s=0} 
\label{eq:371}
\end{equation}

Taking into account that

\begin{equation}
F(s + {{3} \over {2}},s + 1,s + 2, {{5} \over {9}}) = {{\Gamma(s+2)} \over
{\Gamma(1) \  \Gamma(s+1)}} \ \int_0^1  t^s \ \left(1 - {{5t} \over {9}} 
\right)^{-s-{{3} \over {2}}} \ dt
\label{eq:372}
\end{equation}

\noindent we find again a situation of the form

\begin{equation}
F^{\prime}(s + {{3} \over {2}}, s + 1,s + 2, {{3} \over {4}})\vert_{s=0} \  = I_1 +
I_2 + I_3
\label{eq:373}
\end{equation}

\noindent where in this case we have that

\begin{equation}
I_1 = \int_0^1 \left(1 - {{5t} \over {9}}\right)^{-{{3} \over {2}}} \ dt
\label{eq:374}
\end{equation}

\begin{equation}
I_2 = \int_0^1   \left(1 - {{5t} \over {9}}\right)^{-{{3} \over {2}}} 
\ \ln t \ dt
\label{eq:375}
\end{equation}

\begin{equation}
I_3 = -  \int_0^1   \left(1 - {{5t} \over {9}}\right)^{-{{3} \over {2}}} \ 
\ln\left(1 - {{5t} \over {9}}\right) \ dt
\label{eq:376}
\end{equation}

After a tedious but straightforward computation we can write that

\begin{equation}
I_1 = {{9} \over {5}}
\label{eq:377}
\end{equation}

\begin{equation}
I_2 = {{36} \over {5}}  \ln 5 - {{36} \over {5}} \ln 2 -
{{36} \over {5}}  \ln 3
\label{eq:378}
\end{equation}

\begin{equation}
I_3 = - {{54} \over {5}} \ln 2 + {{54} \over {5}} \ln 3 - {{18} \over {5}}
\label{eq:379}
\end{equation}

In doing so $\zeta_r^{\prime}(0)$ corresponds to

\begin{equation}
\zeta_r^{\prime}(0) = 4 \ln 2 + \ln 3 + \ln 5
\label{eq:380}
\end{equation}

\vfill \eject


\begin{thebibliography}{99}

\bibitem{ju}
{G. Junker,} {\it Supersymmetric Methods in Quantum and Statistical Physics}.
Berlin: Springer-Verlag (1996).
\bibitem{ro}
{G. Junker and P. Roy,} Ann. of Phys. (NY) {\bf 256} (1997) 302.
\bibitem{du}
{G. V. Dunne and K. Rao,} Journ. High Enery Phys. {\bf 01} (2000) 019.
\bibitem{mu}
{H. J. W. M\"uller-Kirsten, J. Zhang and Y. Zhang,} 
Journ. High Enery Phys. {\bf 011} (2000) 0111.
\bibitem{kl}
{H. Kleinert,} {\it Paths Integrals in Quantum Mechanics, Statistics and
Polymer Physics}.
Singapore: World Scientific (1990).
\bibitem{ra}
{P. Ramond,} {\it Field Theory: A Modern Primer}.
New York: Addison-Wesley Publishing Company (1992).
\bibitem{rj}
{R. Rajaraman,} {\it Solitons and Instantons}.
Amsterdam: North-Holland (1982).
\bibitem{bo}
{E. B. Bogomol'nyi,} Sov. Journ. Phys. {\bf 24} (1973) 1888.
\bibitem{ha}
{S.W. Hawking,} Commun. Math. Phys. {\bf 55} (1977) 133.
\bibitem{ja}
{M. A. Jafarizadeh and H. Fakhri,}  Phys. Lett. {\bf A230} (1997) 157.
\bibitem{gr}
{I.S. Gradshteyn and I.M. Ryzhik,} {\it Table of Integrals, Series and Products}.
New York: Academic Press (1965). 
\bibitem{ab}
{M. Abramowitz and I.A. Stegun,} {\it Handbook of Mathematical Functions}.
New York: Dover Publications (1970).
\bibitem{co}
{S. Coleman,} {\it Uses of Instantons} in {\it The Whys of Subnuclear Physics}.
Ed. A. Zichichi.
New York: Plenum Press (1979).
\bibitem{ki}
{C. Kittel,} {\it Introduction to Solid State Physics}.
New York: John Wiley and Sons (1959).
\bibitem{ni}
{A.J. Niemi and G.W. Semenoff,}  Phys. Rep. {\bf 135} (1986) 99.


\end{thebibliography}
\end{document}